\documentclass[aps,showpacs,pre,twocolumn]{revtex4-1}

\usepackage[T1]{fontenc}
\usepackage[latin1]{inputenc}
\usepackage{amsmath}
\usepackage{amsthm}
\usepackage{amssymb}
\usepackage{subfigure}	
\usepackage{graphicx}
\usepackage{epstopdf}
\usepackage{color}
\usepackage[colorlinks=true,linkcolor=blue,citecolor=blue]{hyperref}
\usepackage{extramarks}
\usepackage{ae}
\usepackage{color}
\begin{document}
\title{Two-dimensional rogue waves on zero background of the Davey-Stewartson II equation}
\author{Lijuan Guo$^{1}$, Jingsong He$^{2\ast}$, Lihong Wang$^{3}$,  Yi Cheng$^{1}$, D.J. Frantzeskakis$^{4}$, P.G. Kevrekidis$^{5}$ }
\affiliation {$^{1}$ School of Mathematical Sciences, University of Science and Technology of China, Hefei,
Anhui 230026, P.\ R.\ China\\
$^{2}$ Institute for Advanced Study, Shenzhen University, Shenzhen, Guangdong 518060, P.\ R.\ China\\
$^{3}$ School of  Mathematics  and Statistics, Ningbo University,
Ningbo, Zhejiang 315211, P.\ R.\ China \\
$^{4}$ Department of Physics, University of Athens, Panepistimiopolis, Zografos, Athens 15784, Greece \\
$^{5}$ Department of Mathematics and Statistics, University of Massachusetts, Amherst, MA 01003-4515, USA
}
\thanks{Email of corresponding author: hejingsong@szu.edu.cn, jshe@ustc.edu.cn}
%%%%%%%%%%%%%%%%%%%%%%%%%%%%%%%%%%%%%%%%%%%%%%%%

\begin{abstract}

%A novel analytical structure of eigenfunctions of Lax pair associated with zero ``seed''
  A prototypical example of a rogue wave structure in a two-dimensional
  model is presented in the context of the Davey-Stewartson~II (DS~II)
  equation arising in water waves.
%i.e., coefficients of Taylor expansion of an eigenfunction form a
%hierarchy of infinitely many new eigenfunctions,
  The analytical methodology involves a Taylor expansion of an eigenfunction
  of the model's Lax pair which is used  to form a
hierarchy of infinitely many new eigenfunctions. These are used for the construction of
%Using these new eigenfunctions,
two-dimensional (2D) rogue waves (RWs) of the DS~II equation
by the even-fold Darboux transformation (DT). The obtained 2D RWs, which are localized in both
space and time, can be viewed as a 2D analogue of the Peregrine soliton
and are thus natural candidates to describe oceanic RW phenomena,
%, as well as
%RW holes in
as well as ones in 2D fluid systems and water tanks.
%
%may be used
%that is, as $|t|$ and $r'\rightarrow \infty$, its maximum amplitude
%decays as $O(\frac{1}{r'})\sim0$.  Hence, this kind of RWs provides a proper candidate to describe
%for the description of oceanic RW phenomena. In addition, they can also
%describe RW holes in 2D fluid systems and water tanks.
%
%and a two-dimensional analogue of the Peregrine soliton of the NLS equation.
% Meanwhile, the two-dimensional RWs presented here  can also be used to
%describe RW holes in two-dimensional fluid system and water pool.
% Besides, a  rogue wave-lump solution of the DS II is  produced by odd-fold DT,  which  is  not localized in time,  because one lump with a height $\frac{2}{s_0}$ in  this solution is  located
%  at origin of coordinate  as $|t|\rightarrow\infty$.

\end{abstract}

%%%%%%%%%%%%%%%%%%%%%%%%%%%%%%%%%%%%%%%%%%%%%%%%

\maketitle
%\vspace{-0.9cm}

%\noindent {{\bf Keywords}: Davey-Stewartson II equation, Rogue wave, Ocean wave}

%05.45.Yv:soliton;
%02.30.Ik,integrable system;
%03.75.Lm, Tunneling, Josephson effect, Bose¨CEinstein condensates in periodic potentials,
        %solitons, vortices, and topological excitations
%42.65.Tg, Optical solitons; nonlinear guided waves

%%%%%%%%%%%%%%%%%%%%%%%%%%%%%%%%%%%%%%%%%%%%%%%%
%\section{Introduction}

{\it Introduction.} A two-dimensional (2D) rogue wave (RW) 
%in two spatial dimensions (2D)
is a short-lived large amplitude wave, which
is doubly localized in two spatial variables, $x$ and $y$ (as well as in time), 
and its modulus is a rational function.
%The contour line of a 2D RW, above its asymptotic spatial plane (or background), is a closed
%curve which reflects vividly the localization character.
%It is naturally expected that 
Naturally, a RW of a suitable 2D partial differential equation 
%model
can provide a genuine dynamical paradigm of the oceanic RWs
\cite{S2008,annu2008,A2010}, and also be of relevance 
%which occur in a relatively calm sea. Such
%ocean RWs may have a significant impact on loads and responses of ships and
%offshore structures. However, relevant waveforms are of interest
to other areas of physics, including nonlinear optics and atomic
Bose-Einstein condensates~\cite{baronio}. However, any analytical form 
of such a genuine 2D RW has not been reported so far.

A candidate model that may give rise to purely 2D RWs, which in turn may describe
oceanic extreme events, is the Davey-Stewartson~II (DS~II) system \cite{PRSA338}:
\begin{equation}\label{DSII}
\begin{aligned}
&iu_t+u_{xx}-u_{yy}+(2\kappa|u|^2+S)u=0,\\
&S_{xx}+S_{yy}=-4\kappa(|u|^2)_{xx}, \quad \kappa=\pm 1.
\end{aligned}
\end{equation}
In the context of water waves, $u(x,y,t)$ is the amplitude of a surface wave packet, and 
$S(x,y,t)$ is the velocity potential of the mean flow interacting with the surface wave; 
in fact, $u$ is proportional to the amplitude of the first harmonic and $S$ the mean 
(zeroth harmonic) in the slowly-varying envelope expansion.
The fluid velocity potential $\phi$
can then be reconstructed as
$\phi \sim S(x,y,t) + F [u(x,y,t) exp(ikx-i\omega t) + {\rm c.c.}]$,
where c.c. stands for the complex conjugate and
$F=cosh(k(z+h))/cosh(kh)$; $k$ is the wavenumber, $\omega$ the frequency
and $h$ represents the depth.
Finally, parameter $\kappa$ 
sets, in general, the type of nonlinearity: $\kappa=\pm1$ 
corresponds to defocusing and focusing cases; in the water wave problem, $\kappa=-1$.   
This model describes two-dimensional water waves with a weak surface tension
\cite{PRSA338,DR1977,JFM1979} (see also the reviews \cite{mark1981,mark2011} and 
Ref.~\cite{craig} for a rigorous justification).

%The relevance of this system is multifaceted. 
%Firstly, the DS~II equation describes two-dimensional water waves with a weak surface tension
%\cite{PRSA338,DR1977,JFM1979} (see also the reviews \cite{mark1981,mark2011} and 
%Ref.~\cite{craig} for a rigorous justification). In addition, the DS~II equation is
%a natural 

The DS~II equation is the integrable 2D extension of the famous nonlinear Schr\"odinger 
(NLS) equation~\cite{sulem,ablowitz1}. The latter, is known to possess one-dimensional (1D) 
rational RWs, i.e., the fundamental RW in the form of the so-called Peregrine 
soliton~\cite{Peregrine1983,akhmediev1985,akhemediev2009},
as well as other higher-order RWs \cite{akhmediev2011,akhmediev2013,jshe2013, jshe2017},
which have been observed in optical systems and water tank experiments
\cite{akhmediev2010np,akhmediev2011ol,dias2014,kibler2016,chabchoub2011,chabchoub2012,chabchoub2016}.
%Furthermore, 
Since the DS~II equation can be viewed as a nonlocal hyperbolic NLS equation, it should
be mentioned that its local counterpart, i.e., the 2D hyperbolic NLS,
has been 
%shown numerically to
%support RWs
computationally explored in~\cite{oson} in the context of deep
water gravity waves, with the numerical 
results suggesting that RWs may persist in such settings.
%Hence, once 2D RW solutions of the DS~II equation are found,
%they can be viewed as generalizations of the NLS' RWs,
%which can thus model genuine oceanic RWs.

The main focus of the present work is to offer {\it explicit analytical
solutions} for the DS~II RWs, which can be viewed as the prototypical
generalization of Peregrine RWs to genuinely higher dimensional settings.
The ubiquitous relevance of such settings in physical applications
highlights the importance of a systematic toolbox enabling the
identification of such solutions. While the importance of this problem
has been recognized and other mathematically motivated (from integrability
theory) 2D extensions of RWs have been proposed~\cite{kundu}, the
present 2D RW proposal constitutes a prototypical physically relevant
example in view of the above remarks.

Before proceeding, we review 
%it is important to mention 
some key properties of the
DS~II equation. First we note that it is a completely integrable system via the inverse
scattering transform (IST), and particularly by
means of the so-called $\bar{\partial}$-method \cite{JMP25,Fokasbook,book}. In addition, the
DS~II equation, apart from the context of water waves, plays an important role in a variety
of other physical settings, including plasma physics \cite{zakhpr,JPSJ1993,kofane2017},
nonlinear optics \cite{newell,leblond1998,leblond2005}, and
%electromagnetic waves in
ferromagnets \cite{leblond1999}. Regarding its solutions, the defocusing DS~II 
equation ($\kappa=+1$), does not possess lump solutions \cite{physd18}, while 
%, as was rigorously proved in Ref.~\cite{physd18}. On the other hand, 
the focusing DS~II equation ($\kappa=-1$) not only possesses lump solutions
\cite{JMP25,JMP20,Physd1989}, but also line rogue waves \cite{JPA46}
(localized only in one spatial variable), and
lump-line soliton solutions \cite{fokasphysd}. The $N$-lump solution in Ref.~\cite{Physd1989}
displays simple interaction dynamics, namely lumps recover their original velocity and shape
without a phase shift after their interaction. On the other hand, the multi-pole lumps
of the DS~II equation \cite{PLA1997,SIAM2003} feature nontrivial interaction dynamics,
and decay as $r^{-N}~(N\geq2),~r^2=x^2+y^2$. This kind of lumps \cite{SIAM2003} corresponds
to a discrete spectrum whose related eigenfunctions have higher-order poles in
the spectral variable in the framework of IST. The multi-pole lumps were also found in
other 2D integrable equations, such as the Kadomtsev-Petviashvili I (KP I) equation \cite{PRL1997}
and Fokas equation \cite{JPA2017}. Note that the periodic soliton resonance of
the DS~II equation has also been studied in Refs.~\cite{tajiri2001,tajiri2015}.
This broad spectrum of recent activity showcases the wide interest in this
class of models.

Although there exist many results on the solutions of the DS~II equation, the construction of
2D RWs for this equation, is still an open problem, strongly motivated
by (a) modeling analytically 2D RWs in the ocean (or in a large water tank), and
(b) finding a 2D analogue of the Peregrine soliton.
%
%Here it should be pointed out
%that line RWs \cite{JPA46} are not 2D RWs of the DS~II equation, because they
%are not localized in both spatial variables, $x$ and $y$, due to the infinitely long
%line profile of this kind of rational solution.
%
In this Letter, we present genuine 2D RWs of the DS~II equation, decaying
as $O(1/r'^{2})$, for a given time $t$, where prime denotes the moving reference frame
(see below). Unlike higher-order multi-pole lumps \cite{SIAM2003} whose maximum
amplitude approaches to a nonzero constant as $t\rightarrow \pm\infty$,
here we present two kinds of solutions featuring different asymptotic behaviors for large $|t|$:
(i) the maximum value of the first-order RW decays  $\sim 1/r'$ to 0,
which guarantees that this is indeed a RW wave ``appearing from nowhere and disappearing without a
trace''~\cite{akhemediev2009}; (ii) a RW-lump solution, whose maxima consist of a
central peak given by a lump, and an outer ring given by a RW, which eventually reduces
to a lump of constant amplitude
%$\frac{2}{s_0}$
for large $t$.
%a rogue wave-lump solution, which consists of a lump in the center and a RW in outer ring, reduces to a lump possessing a constant amplitude  for large time $|t|$.

{\it Analysis.} In what follows we provide an outline of the methods and give the main results. We start
by recalling the Lax pair and Darboux transformation (DT) for the DS~II equation \cite{matveev,zhoubook,xinkou}. The relevant Lax pair is \cite{matveev,xinkou}:
\begin{equation}\label{lax}
\Psi_y=J\Psi_x+U\Psi,\quad \Psi_t=2J\Psi_{xx}+2U\Psi_x+V\Psi,
\end{equation}
with a constant diagonal matrix $ J=\left(\begin{array}{cc}
i &0\\
0 &-i
\end{array}\right) $,
and two potential matrices:
\begin{equation}
U=\left(\begin{array}{cc}
0 &u\\
v &0
\end{array}\right),
V=\left(\begin{array}{cc}
(w+iQ)/2 &u_{x}-iu_{y}\\
v_{x}+iv_{y} & (w-iQ)/2
\end{array}\right).
\end{equation}
Here, the eigenfunction $\Psi=(\psi,\phi)^T$ ($T$ denotes transpose),
the potentials $u,~v=ku^* \in \mathbb{C}$, and the field $Q=2\kappa|u|^2+S \in \mathbb{R}$,
are functions of the three variables $x, y, t$. As is typically
the case in such integrable models, the DS~II equation is obtained from the
compatibility of the Lax pair, i.e.,  $\Psi_{yt}=\Psi_{ty}$.
Since $S$ and $Q$ are two auxiliary functions in the DS~II and its Lax pair, below
we will focus on the $u$ waveform.

The construction of the $N$-fold DT of the DS~II equation, necessitates
$N$ eigenfunctions
$\Psi_k=(\psi_k,\phi_k)^T~(k=1,2,\ldots,N)$ and $\widetilde{\Psi}_k=(\phi_k^*,\kappa\psi_k^*)^T$,
which are  associated with a given ``seed" solution $u$ and $v$,
of the Lax pair equation~(\ref{lax}). The $N$-th order solution \cite{xinkou}
of the DS~II equation generated by the $N$-fold DT reads:
\begin{equation}\label{uN}
u^{[N]}=u+2i\frac{\delta_2}{\delta_1},
\end{equation}
where  $\delta_1$ and $\delta_2$ are two determinants of $\Psi_k=(\psi_k,\phi_k)^T$ and $\widetilde{\Psi}_k=(\phi_k^*,\kappa\psi_k^*)^T$ (see the Supplemental
Material $S_1$).
The line RWs were constructed in Ref.~\cite{xinkou}. As mentioned above,
the defocusing DS~II model, for $\kappa=1$, has no smooth rational solutions; thus, hereafter,
we focus on the focusing case, and fix $\kappa=-1$.

We are now in a position to construct RW solutions $u$ of the DS~II equation
starting from a zero seed solution, $u=0, v=0, Q=0, w=0$,  of Eq.~(\ref{uN})
by the DT method.
It is crucial to find proper eigenfunctions $\Psi_k$ associated
with the zero seed
solution in the Lax pair equation~(\ref{lax}) in order to find RWs.
Substituting this seed back into Eq.~(\ref{lax}), we get a basic eigenfunction $\Psi$,
namely:
\begin{equation}\label{DSIIfun}
\begin{aligned}
&\psi=\psi(b_1,x,y,t)=a_1\exp[ib_1(x+iy-2b_1t)],\\
&\phi=\phi(b_2,x,y,t)=a_2\exp[ib_2(x-iy+2b_2t)],
\end{aligned}
\end{equation}
where
\begin{equation}
\begin{aligned}
a_1=&\exp{[b_1(s_0+s_1\epsilon+s_2\epsilon^2+\cdots+s_N\epsilon^N)]},\\
 a_2=&\exp{[b_2(s_0+s_1\epsilon+s_2\epsilon^2+\cdots+s_N\epsilon^N)]},
\end{aligned}
\end{equation}
are two overall factors added intentionally in order to introduce more expansion coefficients $s_i$. %It is important to mention that
Importantly, the two components $\psi$ and $\phi$ in the basic eigenfunction $\Psi$ are
independent from each other under the condition of  zero seed solution. Performing a Taylor expansion for the above basic eigenfunction  $\Psi$ at $(\lambda_1, \lambda_2)^{T}$ we obtain:
\begin{equation}\label{taylor}
\begin{aligned}
&\psi(\lambda_1+\epsilon)=\psi^{[0]}+\psi^{[1]}\epsilon+\cdots+\psi^{[N]}\epsilon^{N}+O(\epsilon^{N+1}),\\
&\phi(\lambda_2+\epsilon)=\phi^{[0]}+\phi^{[1]}\epsilon+\cdots+\phi^{[N]}\epsilon^{N}
+O(\epsilon^{N+1}),\\
\end{aligned}
\end{equation}
where $\psi^{[k]}=\frac{1}{k!}\frac{\partial^{k} \psi}{\partial b_1^k}|_{b_1=\lambda_1},
\phi^{[j]}=\frac{1}{j!}\frac{\partial^{j} \phi}{\partial b_2^j}|_{b_2=\lambda_2}(k,j=0,1,2,\ldots,N)$.
By a tedious calculation, we find that $(\psi^{[k]},~\phi^{[j]})^T$ are analytical and
infinitely many eigenfunctions of Lax pair equation~(\ref{lax}) are
associated with
the zero seed solution; this is a novel analytical structure of eigenfunctions
of
the Lax pair for the DS~II equation.
Due to the independence of $\psi$ and $\phi$, these
eigenfunctions are classified in  two categories, i.e.,
$(\psi^{[k]}, \phi^{[k]})^T$ and $(\psi^{[k]}$,
$\phi^{[j]})^T(k\not=j)$. Note that ${\psi}^{[k]}$ contains $k+1$ parameters $s_p$
($p=0,1,2,\ldots,k$) and $\lambda_1$, while ${\phi}^{[j]}$ contains $j+1$ parameters
$s_p$ ($p=0,1,2,\ldots,j$) and $\lambda_2$.
Below, for simplicity, we set $s_0>0$, $s_j=0$ $(j\geq1)$, and
$\lambda_1=\lambda_2=i\lambda$ ($\lambda\in \mathbb{R}$).
%where $\lambda$ is a real constant.
Furthermore, hereafter, all results are discussed
in a moving reference frame, i.e., $x'=x$, $y'=y+4\lambda t$ and $r'=\sqrt{x'^2+y'^2}$,
which is more convenient in order to investigate the properties of the obtained solutions.
%without loss of generality.

To derive rational solutions $u^{[N]}$ of the DS~II equation,
we select $\Psi_k=(\psi_{k},\phi_{k})^T=(\psi^{[2k-1]}, \phi^{[2k-1]})^T (k=1,2,\ldots,N)$
in Eq.~(\ref{uN}). Setting $N=1$ and $\lambda=1$, Eq.~(\ref{uN}) yields a usual first-order
lump, $u^{[1]}_{\rm lump}=-\frac{2s_0e^{2i(2t-y')}}{x'^2+y'^2+s_0^2}$, and its maximum value
$|u^{[1]}_{\rm lumpM}|= \frac{2}{s_0}$.
Setting $N=2$ and $\lambda=1$, Eq.~(\ref{uN}) yields a second-order rational solution
$u^{[2]}$, corresponding to the first-order RW of the DS~II, namely:
\begin{equation}\label{rw}
u^{[1]}_{\rm rw}=\frac{6s_0N_{\rm rw}}{D_{\rm rw}}e^{-2i(2t+y')},
\end{equation}
where
\begin{equation}
\begin{aligned}
N_{\rm rw}=&-x'^4+y'^4-2s_0^2x'^2+4s_0^2y'^2-s_0^4+\\
&12it(x'^2+y'^2+s_0^2),\\
D_{\rm rw}=&(x'^2+y'^2+s_0^2)^3+12s_0^2(x'^2+s_0^2)y'^2+144s_0^2t^2.\nonumber
\end{aligned}
\end{equation}
The solution $u^{[1]}_{\rm rw}$ is a smooth and nonsingular RW as $s_0>0$ on zero background,
with the following properties:
%
%\begin{itemize}
%
%\item

$\bullet$ The modulus $|u^{[1]}_{\rm rw}|$, which is an  even function,
%namely $|u^{[1]}_{\rm rw}(x',y',t)|=|u^{[1]}_{\rm rw}(-x',-y',-t)|$,
decays algebraically like $O(\frac{1}{r'^2})$ for any given time $t$.
%
%\item

\begin{figure}[!htbp]
\centering
\raisebox{16 ex}{}\subfigure[$t=0$]{\includegraphics[height=2.5cm,width=2.8cm]{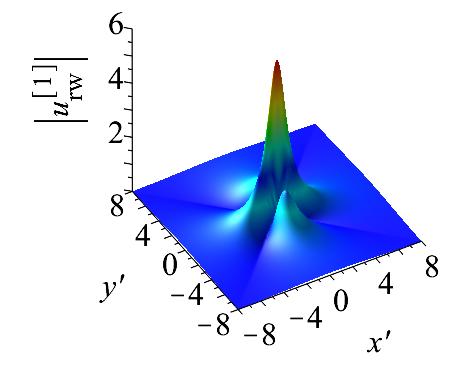}}
\raisebox{12 ex}{}\subfigure[$t=1$]{\includegraphics[height=2.5cm,width=2.8cm]{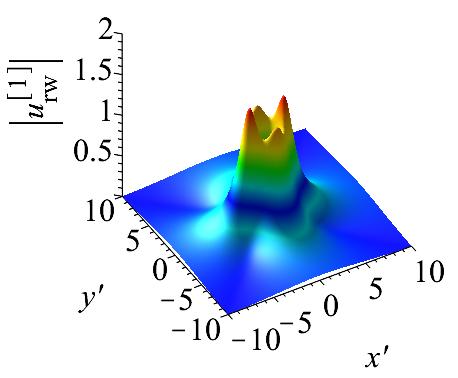}}
\raisebox{12 ex}{}\subfigure[$t=200$]{\includegraphics[height=2.5cm,width=2.8cm]{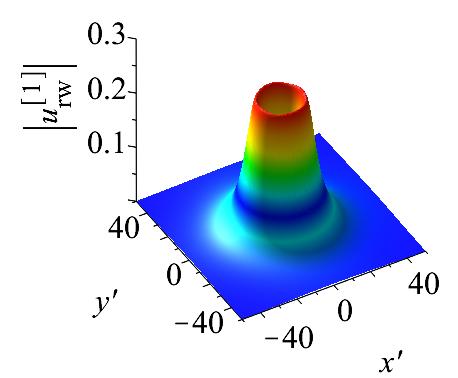}}\\
\raisebox{19 ex}{}\subfigure[$t=0$]{\includegraphics[height=2.5cm,width=2.5cm]{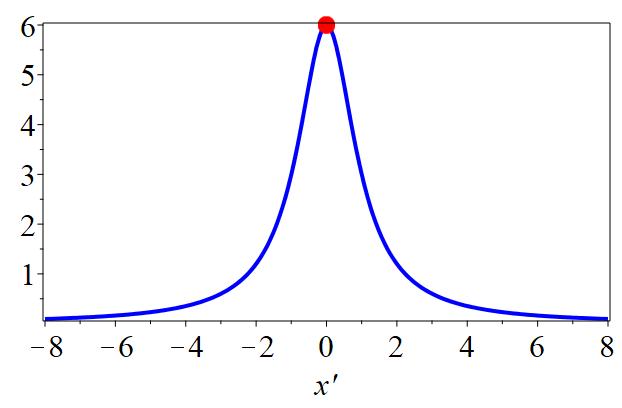}}\quad
\raisebox{19 ex}{}\subfigure[$t=1$]{\includegraphics[height=2.5cm,width=2.5cm]{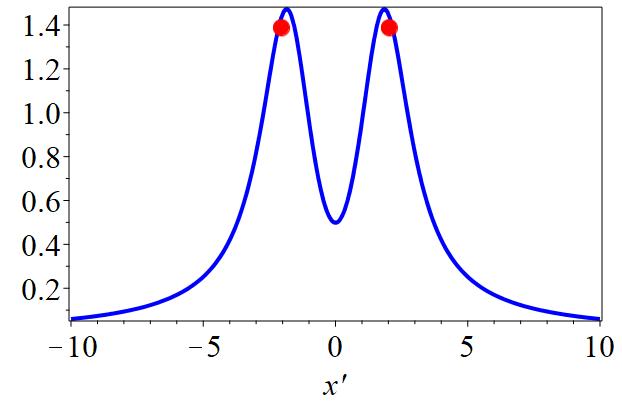}}\quad
\raisebox{19 ex}{}\subfigure[$t=200$]{\includegraphics[height=2.5cm,width=2.5cm]{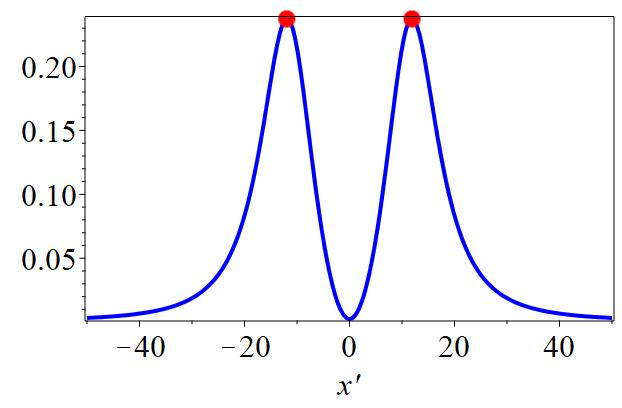}}\quad
\caption{(Color online) Top panels: Profiles of the first-order RW $|u^{[1]}_{\rm rw}|$  in the ($x',y'$)-
plane with parameters $s_0=1$ at $t=0$, $1$, $200$, respectively.
Bottom panels: The corresponding cross-sectional profiles of the top panels along $x'$-axis.  The red solid circles are plotted using
the exact formulation $\frac{6}{s_0}$ for $t=0$, and the approximate formulas $r'$ and
$|u^{[1]}_{\rm rwcircle}|$ for $t\not=0$.
%{\color{red}{\bf PGK: Lijuan, I am not sure that both the top and second panel row
 % is needed. If the top is kept the z axis needs label. If the 2nd
  %row is kept, the colorbar is needed. I am also perplexed
  %about the bottom panel vs. the middle row (I can explain why) + the
 % bottom needs y-labels too. IMPORTANTLY, same things apply to all figures !}}
}\label{newrw}
\end{figure}

$\bullet$ When $r'=\sqrt{x'^2+y'^2}=0$, the RW solution $u^{[1]}_{\rm rw}$ reduces to:
\begin{equation}
u^{[1]}_{\rm rwc}=\frac{(-6s_0^3+72is_0t)e^{-4it}}{s_0^4+144t^2}.
\end{equation}
Notice that $|u^{[1]}_{\rm rwc}|$ reaches the maximum value $\frac{6}{s_0}$ at $t=0$ (see
Figs.~\ref{newrw}(a,d)), while it obtains a (local) minimum value
thereafter,
and finally decays like $O(\frac{s_0}{2t})\sim0$
as $|t|\rightarrow \infty$.
%
%\item

$\bullet$
When $r'\gg0$, the maxima of $|u^{[1]}_{\rm rw}|$ form a rectangular curve of
different height at an intermediate stage (Figs.~\ref{newrw}(b,e)),
which eventually morphs into a circle (i.e., a nearly radial
pattern) at the final stage of the
evolution (Figs.~\ref{newrw}(c,f)). At its maximum, the radius of this circle
is given by:
  \begin{equation}
  r'=\sqrt{x'^2+y'^2}\sim(6s_0\sqrt{2}|t|)^{1/3}
  %{\frac{1}{3}},
  %\quad |t|\rightarrow\infty,
\label{asympt}
  %  \nonumber
  \end{equation}
  as $|t|\rightarrow\infty$, while the corresponding %wavefunction
  intensity
  \begin{equation}
  |u^{[1]}_{\rm rwcircle}|^2 \sim \frac{4}{(3s_0|t|)^{\frac{2}{3}}}\sim\frac{8}{(r')^2}\sim 0, %\text{as}\ |t|\rightarrow\infty,  r'\rightarrow\infty.
%  \nonumber
  \end{equation}
as $|t|\rightarrow\infty$, $r'\rightarrow\infty$.
%Notice that for $r'\gg0$, the intensity maximum of $|u^{[1]}_{\rm rw}|$ does not correspond to $t=0$
%(see Fig.~\ref{newrw}(a,{\color{red}c})).
%\end{itemize}

The
%above analysis on the
symmetry and extreme values of the intensity
with $s_0=1$ can be verified by the snapshots shown in Fig.~\ref{newrw}, for $t=0,1,200$.
Note that only profiles for $t\geq0$ are shown
because $|u^{[1]}_{\rm rw}|$ is an even function of $t$.
The red solid circles denote the maxima of $|u^{[1]}_{\rm rw}|$, which are plotted
by using the exact formulation $\frac{6}{s_0}$ for $t=0$, and the approximate formulas $r'$ and
$|u^{[1]}_{\rm rwcircle}|$ for $t\not=0$. There exists a
deviation between the red solid circles and the maxima in Fig.~\ref{newrw}(e) because time
$t$ is too small for the asymptotics to be valid. This deviation disappears in
Fig.~\ref{newrw}(f) for (sufficiently) large time $t=200$, which is well
%an excellent agreement with the analytical results.
within the asymptotic regime of Eq.~(\ref{asympt}).

An animation \cite{supplemental} is provided in  the Supplementary Material
%by
%using analytical formula $|u^{[1]}_{\rm rw}|$  to
showing the
dynamical evolution of the first order RW, which can be summarized as follows.
At the early stage of the evolution, for large and negative $t$,
$|u^{[1]}_{\rm rw}|$
appears from the background as a wide
circle of low intensity;
gradually, it converges to a rectangular column with four maxima on its top at the
intermediate stage, and then a large peak at $t=0$.
Next, the RW follows the reverse path, initially dispersing and eventually
reverting back to a nearly radial form for large and positive $t$.
%Next, $|u^{[1]}_{\rm rw}|$ disperses
%gradually as a rectangular column with four maxima on its top at the intermedia%te stage,
%then it becomes a large circle possessing a very tiny height, and eventually de%cays to
%the background at the final stage of the evolution, as $t\rightarrow +\infty$.
%Moreover, the amplitude of $u^{[1]}_{\rm rw}$ decays to a small height $0.24$
%at $t=200$, and it approaches zero as $t\rightarrow +\infty$. Hence, Fig.~\ref{%newrw}
%and animation \cite{supplemental} indeed support the argument that $u^{[1]}_{\rm rw}$ is
%a RW on zero background, i.e., $u^{[1]}_{\rm rw}$ is a short-lived large amplitude wave, which is doubly localized in the two spatial variables $x'$ and $y'$.

 %%%%%%%%%%%%%%%%%%%%%%%%%%%%%%%%%%%%%%%%%%%% the first order rw

Here it should be pointed out that the first order RW $u^{[1]}_{\rm rw}$ is very different from
the multi-pole lump \cite{SIAM2003} of the DS~II, because the latter has a non-vanishing amplitude
%$\frac{2}{\rho}$
as $|t|\rightarrow \infty$. In addition, this RW solution is naturally  also different from
the line RW \cite{JPA46} of the DS~II, since the latter is not doubly localized in both
$x$ and $y$, but only in one of the two (in the other, it features a line
profile).
%, as it features an infinitely long line profile in the ($x,y$)-plane.

Extending this approach to $N=3$, Eq.~(\ref{uN}) yields a third-order rational solution, which is an unprecedented (to the best of our knowledge)
RW-lump
solution $u_{\rm rw-lump}$ of the DS~II equation. The explicit expression of $u_{\rm rw-lump}$
\cite{supplemental} is given in the Supplementary Material $S_2$.
 % {\color{red}   {\bf PGK:
 % A separate file needs to be created with this and also
 % the other items indicated here that they should
 % be placed in the supplement}}.
Here, we describe its
principal dynamical
properties.
%\begin{itemize}
%
%\item

$\bullet$ The modulus $|u_{\rm rw-lump}|$, which is an  even function,
%  namely $|u_{\rm rwlump}(x',y',t)|=|u_{\rm rwlump}(-x',-y',-t)|$,
decays algebraically like $O(\frac{1}{r'^2})$ for any given time $t$.
%
% \item

\begin{figure}[!htbp]
\centering
\raisebox{19 ex}{}\subfigure[$t=0$]{\includegraphics[height=2.5cm,width=2.8cm]{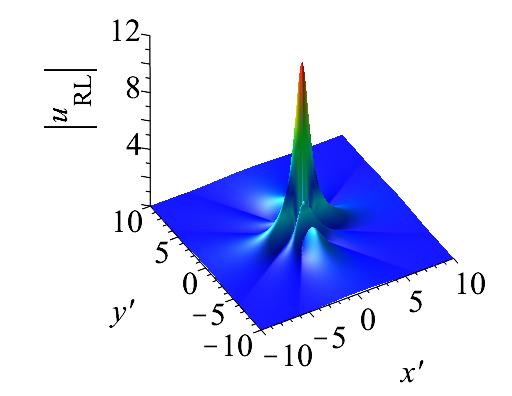}}
\raisebox{19 ex}{}\subfigure[$t=2$]{\includegraphics[height=2.5cm,width=2.8cm]{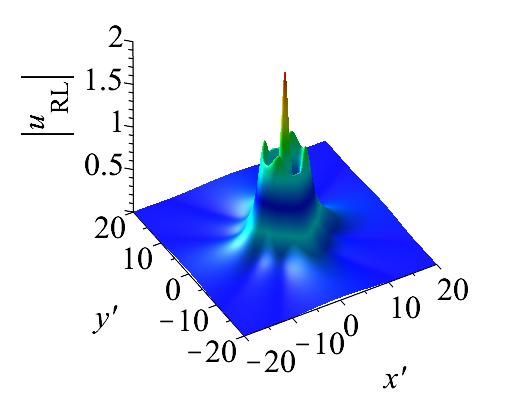}}
\subfigure[$t=200$]{\includegraphics[height=2.5cm,width=2.8cm]{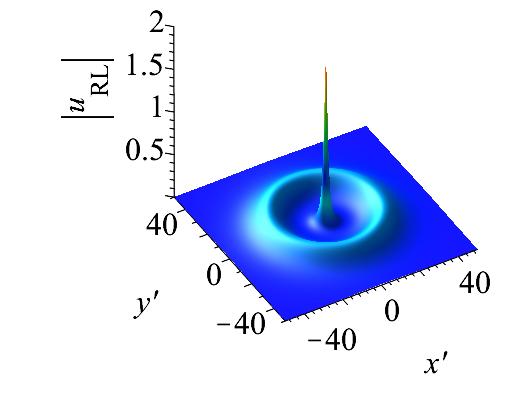}}\\
\raisebox{19 ex}{}\subfigure[$t=0$]{\includegraphics[height=2.5cm,width=2.8cm]{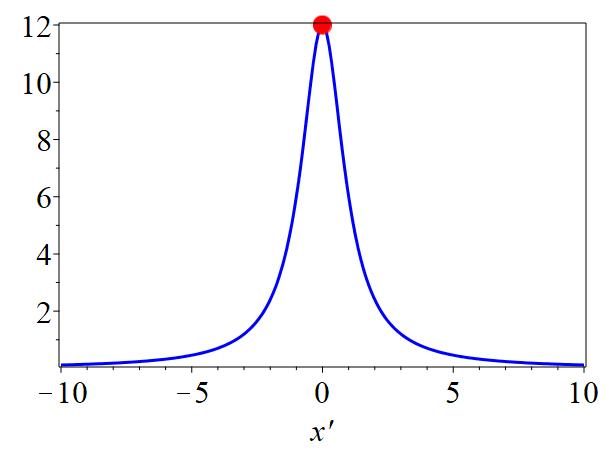}}
\raisebox{19 ex}{}\subfigure[$t=2$]{\includegraphics[height=2.5cm,width=2.8cm]{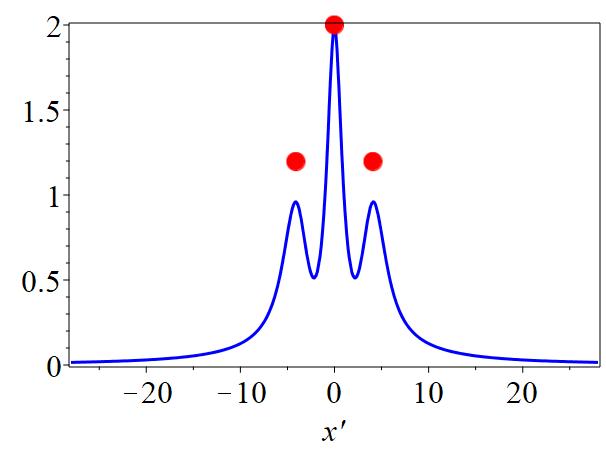}}
\raisebox{19 ex}{}\subfigure[$t=200$]{\includegraphics[height=2.5cm,width=2.8cm]{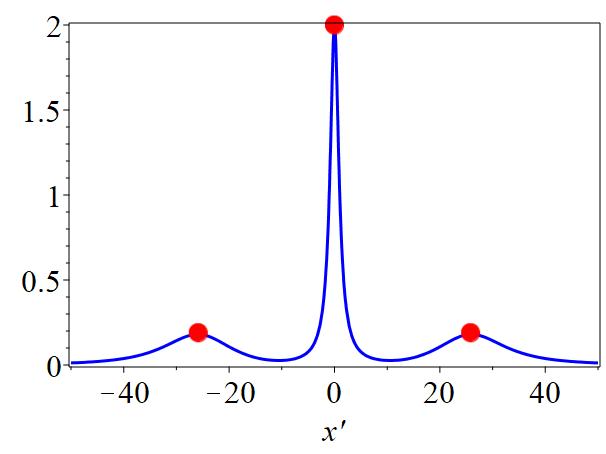}}
\caption{(Color online) Top panels: Profiles of the rogue wave-lump $|u_{\rm rw-lump}|$
  ($|u_{\rm {\tiny{RL}}}|$ in this figure for convenience)
  in the ($x',y'$)-plane with parameters $s_0=\lambda=1$ at $t=0$,
  $2$, $200$, respectively.
Bottom panels: The corresponding cross-sectional profiles of the top panels along $x'$-axis.  The red solid circles are plotted using the
exact formulation $\frac{12}{s_0}$ for $t=0$, and the approximate formulas $\frac{2}{s_0}$($r'=0$)  and  $ \frac{2\sqrt{6}}{r'}$ ($r'\not=0$) for $t\not=0$.}
  \label{newrwlump}
\end{figure}

$\bullet$ When $r'=0$, the
%rogue wave-lump
solution $u_{\rm rw-lump}$ reduces to
  \begin{equation}
  u_{\rm rw-lumpc}=\frac{12(-s_0^4+60t^2+10is_0^2t)e^{-4i\lambda^2t}}{s_0(s_0^4+360t^2+60is_0^2t)}.
  \end{equation}
The modulus $|u_{\rm rw-lumpc}|$ reaches the maximum value
$\frac{12}{s_0}$ at $t=0$ (Figs.~\ref{newrwlump}(a,d)), as a result
of the interaction of the RW and the lump.
Then, it returns to
 maximum value of the first-order lump, i.e., $|u^{[1]}_{\rm lumpM}|\sim \frac{2}{s_0}$,
 as  the RW disappears for $|t|\rightarrow \infty$.
 %
 %\item

 $\bullet$ For $r'\gg0$, maxima of $|u_{\rm rw-lump}|$ form a rectangular curve
 at the intermediate stage (Figs.~\ref{newrwlump}(b,e)), which subsequently
 reverts to a radial outgoing structure at the final stage (Figs. \ref{newrwlump}(c,f)) of the evolution --in addition to the persisting lump at the center.
At its maximum, the radius of this circle is given by:
      \begin{equation}
      r'=\sqrt{x'^2+y'^2}\sim \left(288s_0|t|^2\right)^{1/5},
      %\quad \it {|t|\rightarrow \infty},
\label{asympt2}
 %     \nonumber
      \end{equation}
      as $|t|\rightarrow \infty$, while the intensity %of the wavefunction
      at this
      %(asymptotic)
      circle is:
      \begin{equation}
      \begin{aligned}
     |u_{\rm rw-lump-circle}|^2 \sim\frac{{2}{\sqrt{3}}^{\frac{2}{5}}}{({s_0|t|^2})^{\frac{2}{5}}}\sim\frac{24}{(r')^2},
%     \quad {|t|\rightarrow \infty},    {|r'|\rightarrow \infty}.
     %    \nonumber
     \end{aligned}
     \end{equation}
%\end{itemize}
%
as $|t|\rightarrow \infty$, $r'\rightarrow \infty$.

It is worth observing that the maxima of $|u_{\rm rw-lump}|$ can be decomposed into
$|u^{[1]}_{\rm lumpM}|+ |u_{\rm rw-lump-circle}| \sim \frac{2}{s_0}+  \frac{2\sqrt{6}}{r'}$
for  $|t|\gg 0$, supporting the asymptotic decomposition into a lump and a RW
as discussed above.
Fig.~\ref{newrwlump} presents the relevant features through the
snapshots of different times ($t=0$,  $2$, $200$), incorporating where
possible predictions of the analytical formula, such as the location
of the RW circle for large $t$.
%Therefore, maxima of the rogue wave-lump consist of a central peak given by a lump
%and an outer ring given by a RW.
% as $t\gg 0$.
%Note that the rogue wave-lump $u_{\rm rw-lump}$ reduces finally to
%a lump approaching the constant amplitude $\frac{2}{s_0}$ for large time $|t|$.
%The above features of the rogue wave-lump solution $u_{\rm rw-lump}$,
%are depicted in Fig.~\ref{newrwlump}, where three profiles
%of its modulus $|u_{\rm rwlump}|$ are plotted, for $t=0,2,200$, with $\lambda=1$ and $s_0=1$.
%The red solid circles denote the maxima of $|u_{\rm rw-lump}|$, which are plotted by using
%the exact formula $\frac{12}{s_0}$ for $t=0$, and the approximate formulas $\frac{2}{s_0}$, $r'$
%and $|u_{\rm rw-lump-circle}|$ for $t\not=0$.
%As before, the deviation, for small time $t$, between red solid circles and maxima in
%Fig.~\ref{newrwlump}(h), disappears in Fig.~\ref{newrwlump}(i) for large time, $t=200$.
%The red dashed circle in Fig.~\ref{newrwlump}(f) is plotted by using approximate radius $r'$ of
%maxima, which is in an excellent agreement with the Cyan circle (i.e, the positions of maxima)
%given by density plot of $|u_{\rm rw-lump}|$.
The animation provided in~\cite{supplemental}  shows the
dynamical evolution of the RW-lump. One can discern that:
(a) a lump in the center always exists; (b) a RW appears as an outer ring
from the background
at an early stage, then converges gradually to the origin of coordinate as a large peak,
(c) later, the RW is dispersed to a circle, and finally disappears into the background again as $|t| \rightarrow \infty$.

%%%%%%%%%%%%%%%%%%%%%%%% rw lump

\begin{figure}[!htbp]
\centering
\raisebox{19 ex}{}\subfigure[$t=0$]{\includegraphics[height=2.5cm,width=2.5cm]{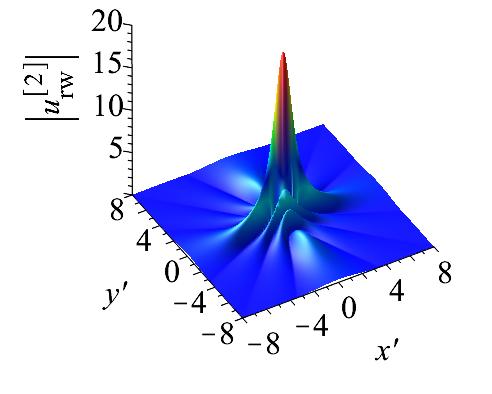}}\quad
\raisebox{19 ex}{}\subfigure[$t=100$]{\includegraphics[height=2.5cm,width=2.5cm]{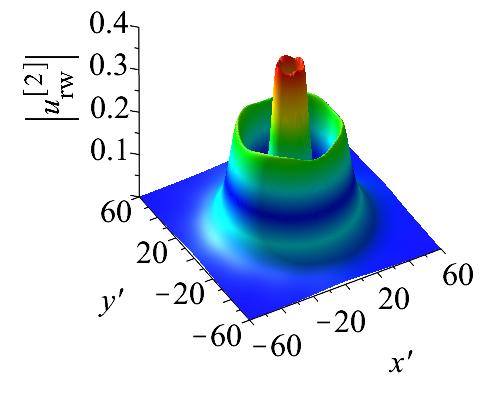}}\quad
\raisebox{19 ex}{}\subfigure[$t=1000$]{\includegraphics[height=2.5cm,width=2.5cm]{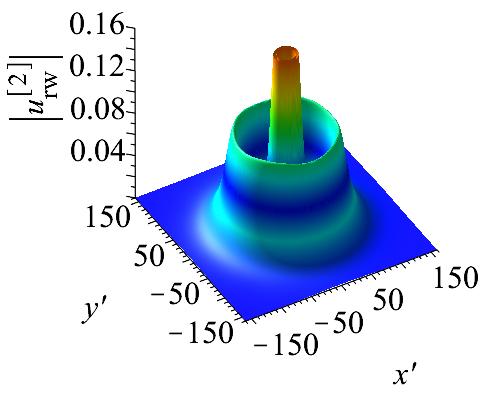}}\\
%\raisebox{19 ex}{}\subfigure[$t=0$]{\includegraphics[height=2.5cm,width=2.5cm]{2rw01.jpg}}\quad
%\raisebox{19 ex}{}\subfigure[$t=100$]{\includegraphics[height=2.5cm,width=2.5cm]{2rw100.jpg}}\quad
%\raisebox{19 ex}{}\subfigure[$t=1000$]{\includegraphics[height=2.5cm,width=2.5cm]{2rw1000.jpg}}\\
\raisebox{19 ex}{}\subfigure[$t=0$]{\includegraphics[height=2.5cm,width=2.5cm]{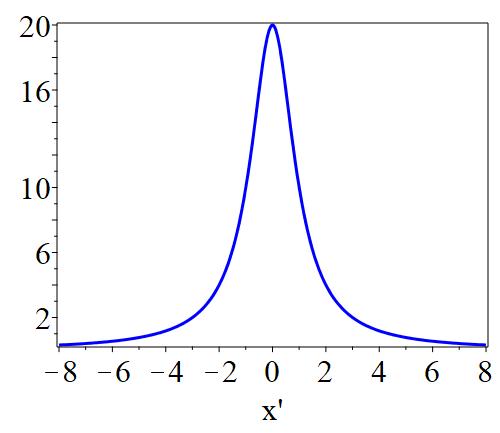}}\quad
\raisebox{19 ex}{}\subfigure[$t=100$]{\includegraphics[height=2.5cm,width=2.5cm]{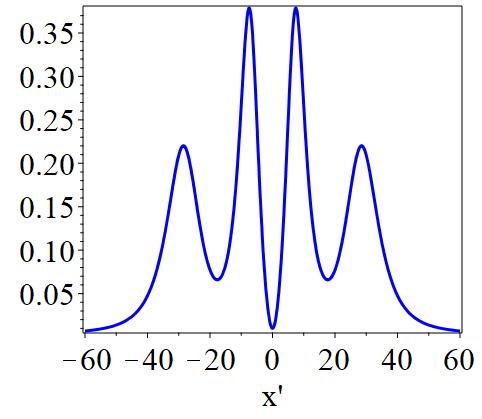}}\quad
\raisebox{19 ex}{}\subfigure[$t=1000$]{\includegraphics[height=2.5cm,width=2.5cm]{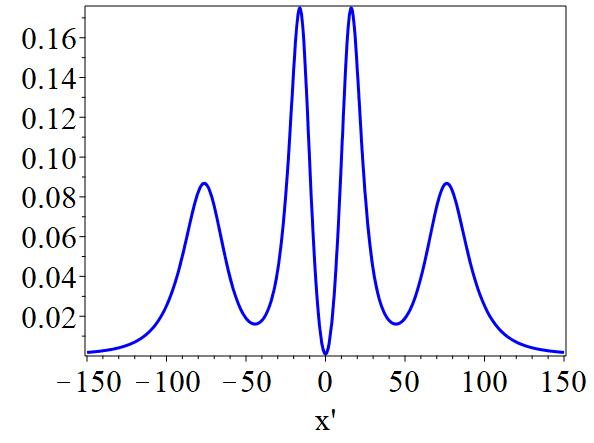}}\\
\caption{(Color online)
Top panels: Profiles of the second-order RW $|u^{[2]}_{\rm rw}|$ in the ($x',y'$)-plane, for
$s_0=\lambda=1$ at $t=0$, $100$, $1000$, respectively.
Bottom panels:
The corresponding cross-sectional profiles of top panels along $x'$-axis.}
\label{2rw}
\end{figure}

Higher-order RWs and RW-lumps  of the DS~II equation can be constructed by
even-fold and odd-fold DT, respectively. For instance,
%To show the applicability of this method clearly,
a second-order RW $|u^{[2]}_{\rm rw}|$, which is constructed by setting $N=4$ in Eq.~(\ref{uN}),
is plotted  in Fig.~\ref{2rw} (see also an animation \cite{supplemental}) with $s_0=\lambda=1$.
Both Fig.~\ref{2rw} and the animation demonstrate again the appearance,
convergence (to $r'=0$),
and then dispersion, and disappearance of the RW. Compared with $u^{[1]}_{\rm rw}$,
the second-order RW has (i) higher amplitude, and (ii) two rings of intensity
maxima for large time $t$.
%Note that, as the modulus of RW solutions is even, the profiles of reversed evolution of
%the RWs for $-\infty<t \le 0$ are also shown in Figs.~\ref{newrw}(g,h,i) and
%Figs.~\ref{2rw}(g,h,i). It is seen that, for $t<<0$, a rogue hole forms at the center
%at the early stage of the evolution  and then gradually becomes a rogue wave peak at $t=0$.
%In the others, there exists remarkable two-dimensional RW holes in RWs during their evolution
%process.
Between them, these harbor a 2D RW hole (see Fig.\ref{2rw}(e,f)) during the
evolution process. It is
interesting to
notice that a 1D RW hole of the NLS equation has been experimentally observed
in a water tank \cite{roguehole}.
%Thus, the 2D RW holes presented here
%can be thought as generalizations thereof in the 2D realm.
%can also be used to
%describe RW holes in 2D settings.
%fluid systems and water.

%%%%%%%%%%%%%%%%%%%%%%%%%%%%%%%%%%second order rw

{\it Conclusions.} We have reported an unprecedented analytical structure of
eigenfunctions of the Lax pair associated with the zero seed solution for the DS~II equation.
%--i.e., the coefficients
%of Taylor expansion of an eigenfunction form a hierarchy of infinitely many new eigenfunctions.
Substituting  these new eigenfunctions into the even-fold DT, we found genuine 2D RWs of the
DS~II equation. We have thus addressed the long-standing problem of the construction
of RWs on zero background for DS~II, and provided a proper candidate to describe fluid RWs
by means of a canonical 2D generalization
of the Peregrine soliton of the 1D NLS equation.
The RWs were localized both in space and time, that is, as $|t|$ and $r'\rightarrow \infty$, their
maximum amplitude decayed to $0$ as $1/r'$.
%$O(\frac{1}{r'})\sim0$.
%Unlike RWs, the lumps of the DSII wrew
%just localized in space which  have non-vanished peaks as  $|t|\rightarrow\infty$.
As a byproduct, RW-lump solutions of the DS~II have been shown
to be generated by odd-fold DT,
which is localized in space only, because one-lump of this solution located at the origin
%of coordinate
approaches  $\frac{2}{s_0}$  as $|t|\rightarrow\infty$. Multi-ring RWs
(obtained through higher-order expansions)
also exist and feature RW holes between them.
%The 2D RWs of the DSII
%can also be used to describe RW holes in 2D fluid systems.

%Although our analytical findings are difficult to be verified in the open ocean,
%they
We expect that these findings may motivate research efforts towards
the generation of 2D RWs in large water tanks
and optical systems, in a multi-dimensional extension of recent 1D
experimental efforts, and paving new directions for RW research.
Moreover, much like line solitons of the
KP~II equation, which have been used to explain shallow water wave patterns \cite{mark2012pre},
the RWs of the DS~II can also be used for similar studies on ocean waves
--since both KP~II and DS~II are pertinent to shallow water with weak surface tension,
and DS~II can be derived from KP~II \cite{mark2011}.

%Full details of this paper  will be published separately. Moreover, RWs on non-zero background of the DS II and DS I and other two-dimensional integrable systems were undertaken.

%%%%%%%%%%%%%%%%%%%%%%%%%%%%%%%%%%%%%%%%%%%%%%%%%%%%%%%%%%%%%%%%%%%%
\textbf{Acknowledgments.}
This work was supported by the National Natural Science Foundation of China (Grants 11671219 and 11871446) and
the Natural Science Foundation of Zhejiang Province (Grants LZ19A010001 and LSY19A010002). P.G.K. and D.J.F. acknowledge that this work was made possible by NPRP Grant No.8-764-1-160 from Qatar National Research Fund (a member of Qatar Foundation).
%The findings achieved herein are solely the responsibility of the authors.

%%%%%%%%%%%%%%%%%%%%%%%%%%%%%%%%%%%%%%%%%%%

%%%%%%%%%%%%%%%%%%%%%%%%%%%%%%%%%%%%%%%%%%%

\end{document}